\documentclass[12pt]{iopart}

\usepackage{graphicx,graphics,color,epsfig}
\usepackage{epstopdf}
\usepackage{bm}

\begin{document}

\title{Origin of pressure induced second superconducting dome in $A_y$Fe$_{2-x}$Se$_2$ [$A$=K, (Tl,Rb)]}

\author{Tanmoy Das$^1$, and A. V. Balatsky$^{1,2,3}$}
\address{$^1$Theoretical Division, Los Alamos National Laboratory, Los Alamos, New Mexico 87545 USA.\\
$^2$Center for Integrated Nanotechnology, Los Alamos National Laboratory, Los Alamos, New Mexico 87545 USA.
$3$ NORDITA, Roslagstullsbacken 23, 106 91 Stockholm, Sweden}

\ead{tnmydas@gmail.com}

\date{\today}

\begin{abstract}
Recent observation of pressure induced second superconducting phase in $A_y$Fe$_{2-x}$Se$_2$ [$A$=K, (Tl,Rb)] calls for the models of superconductivity that are rich enough to allow for multiple superconducting phases. We propose the model where pressure induces renormalization of band parameters in such a way that it leads to changes in Fermi surface topology even for a fixed electron number. We develop low-energy effective model, derived from first-principles band-structure calculation at finite pressure, to suggest the phase assignment where low pressure superconducting state with no hole pocket at $\Gamma$ point  is a nodeless $d$-wave state. It evolves into a $s^{\pm}$ state at higher pressure when the Fermi surface topology changes and hole pocket appears. We analyze the pairing interactions using five band tight binding fitted band structure and find strong pairing strength dependents on pressure. We also evaluate the energy and momentum dependence of neutron spin resonances in each of the phases as verifiable predictions of our proposal.
\end{abstract}

\maketitle

\tableofcontents

\section{Introduction}

High$-T_c$ superconductivity often occurs when the system is driven from its pristine phase to the verge of magnetic quantum critical point via external parameters such as chemical doping, magnetic field or pressure ($P$) in most of the cuprates, heavy fermions, pnictides, and organic superconductors. However, several recent breakthrough discoveries of a second superconducting (SC) dome $-$ completely isolated or slightly connected to the first SC dome $-$ without the intervention of any competing order as a function of $P$ in several families of high-$T_c$ superconductors,\cite{HF,cuprate,KFe2Se2P} and/ or extreme chemical doping in KFe$_2$As$_2$,\cite{KFe2As2} LaFeAsO$_{1-2}$H$_x$\cite{Hosono} or strain\cite{SrFeAs_S} have questioned this expectation. Much higher optimal $T_c$ value than that of the first dome, as well as a substantial increase in electronic mass with $P$ as obtained in iron-based compounds suggest an interesting and exotic phenomena of superconductivity along this tuning axis. It is notable that in stochiometric SrFe$_2$As$_2$, a SC dome appears both as a function of $P$,\cite{SrFeAs_P} as well as crystallographic strain\cite{SrFeAs_S} near the quantum critical point of spin-density wave, as often observed in other high-$T_c$ superconductors, further supporting the notion that the $P$ induced superconductivity is unconventional. In this paper, we present a model that allows us to capture the onset of the second SC phase with different pairing symmetry as a function of $P$.

Our approach is based on the weak or intermediate coupling scenario in which the shape of the Fermi surface (FS) topology plays a key role in creating pairing instability at the `hot-spot'.\cite{Scalapino} For such case Cooper pairing arises from repulsive interaction with sign-reversal pairing symmetry constrained by the FS topology and crystal symmetries. This theoretical framework consistently describes $d-$wave pairing in cuprates,\cite{Scalapino} and Ce-based heavy fermions,\cite{Chubukov} $s^{\pm}$-pairing in iron-pnictides and chalcogenides,\cite{Mazin} and nodeless $d$-wave in $A_y$Fe$_{2-x}$Se$_2$ [$A$=K,Cs, Rb,(Tl,Rb), (Tl,K)] families.\cite{Maier,DHLee,Das}

\begin{figure}[top]
\centering
\rotatebox[origin=c]{0}{\includegraphics[width=0.9\columnwidth]{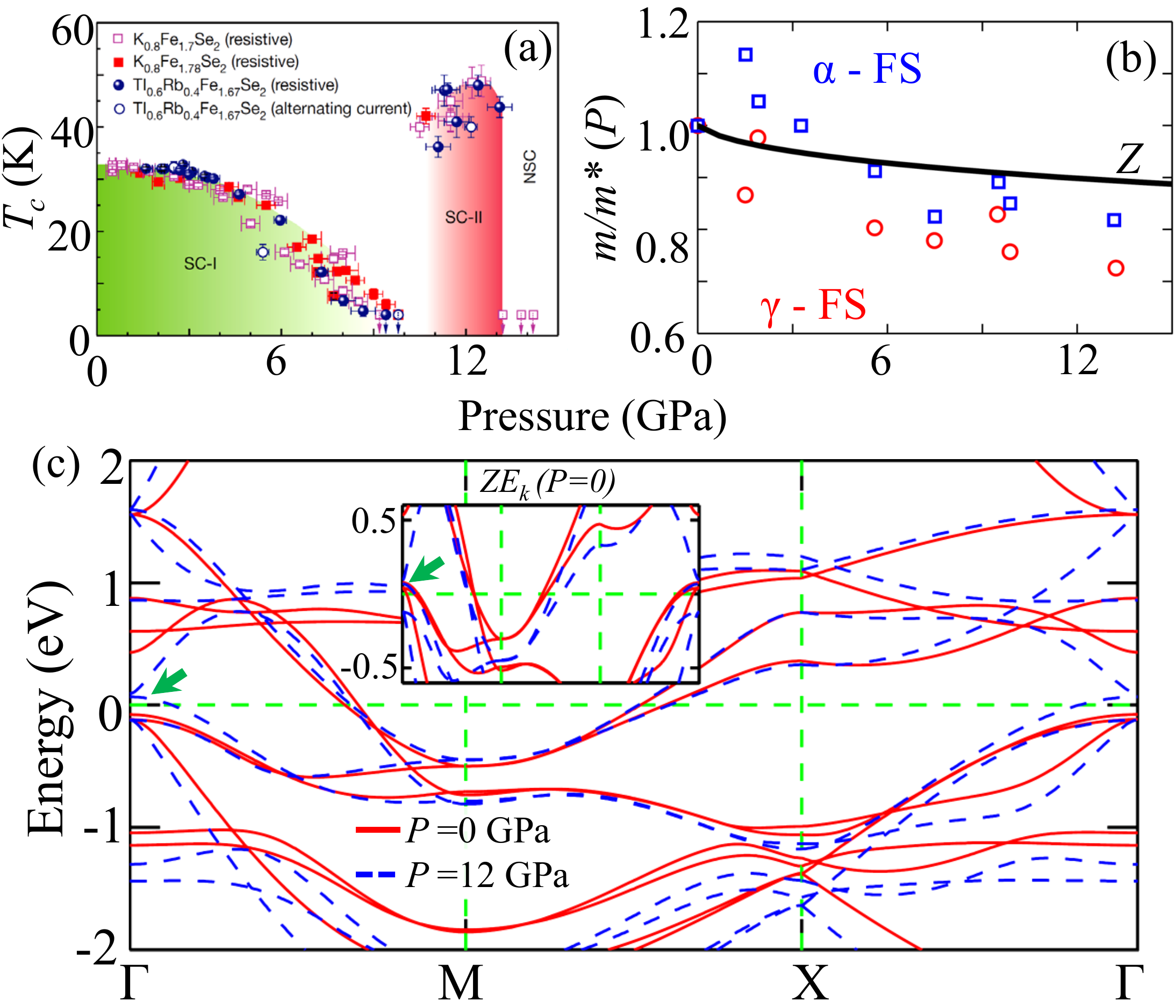}}
\caption{(a) Experimental phase diagram of $T_c$ vs. $P$ for several $A_y$Fe$_{2-x}$Se$_2$ systems, reproduced from Ref.~\cite{KFe2Se2P}. (b) Mass renormalization for hole-pocket ($\alpha$-FS) and electron pocket ($\beta$-FS) for a related compound BaFe$_2$As$_2$, deduced from quantum oscillation measurements (symbols).\cite{massP} The experimental data are normalized to their values at $P=0$. Theoretical results of mass renormaliztion $Z=m/m^*$ for $A_y$Fe$_{2-x}$Se$_2$ compound (see text) plotted in solid line. (c) {\it Ab-inito} band structure of KFe$_2$Se$_2$ at two representative $P$s using TB-LMTO method (see text). {\it Inset:} Band structure at $P=0$ (solid line) is multiplied by $Z=0.9$ as $E_{\bm k}=Z\xi_{\bm k}-E_F$ (with a different Fermi energy to account for the constant electron count). The renormalized band qualitatively reproduces the FS topology obtain at $P$=12~GPa (dashed line), affirming the fact that $P$ increases mass in this system as seen in experiment in (b). Green arrow points to the hole pocket that emerges at high $P$.
} \label{fig1}
\end{figure}

Here, we focus on the latter family in which a second isolated dome is observed, as shown Fig.~\ref{fig1}(a) (reproduced from Ref.~\cite{massP}). In an analogous 122 compound BaFe$_2$As$_2$, it is established that the effective mass, $m^*$, gradually increases as a function of $P$, see the plot of $1/m^*$ in Fig.~\ref{fig1}(b). Constrained by these experimental facts we postulate that uniform $P$ renormalizes the bands in a way that goes beyond standard Fermi-liquid (FL) behavior, as also shown in first-principles calculations,\cite{FSPLDA} and thereby affects the FS topology. In $A_y$Fe$_{2-x}$Se$_2$ systems, both first principles calculation and angle-resolved photoemission spectroscopy (ARPES) have demonstrated that a hole-pocket lies slightly below the Fermi level ($E_F$) at ambient $P$. With band renormalization this hole pocket appears on the FS above a critical $P$, and the overall FS topology changes from only electron-pockets at $P=0$ to the coexisting electron and hole pocket as in pnictides. This topological FS transition induces a crossover from FS nesting along diagonal ${\bm Q}_1\rightarrow (\pi,\pi)$ direction to along zone direction ${\bm Q}_2\rightarrow(\pi,0)/(0,\pi)$ which makes a pairing symmetry transition from nodeless $d$-wave to $s^{\pm}$-pairing.

\section{First-principles band structure}
We explore these postulates via first-principles band structure and pairing eigenvalue calculations within random-phase approximation (RPA). The first-principles calculation is performed for KFe$_2$Se$_2$ within the atomic sphere approximation by using the tight-binding-linearized muffin-tin orbital (TB-LMTO) code.\cite{LMTO,LMTO1} For a given $P$, the uniform volume contraction is evaluated using Birch-Murnaghan equation of state\cite{Birch} formalism (given below). The key here is to optimize the Se atomic position with respect to the Fe plane by minimizing total energy at each $P$. We find $z_{Se}=0.3452c$ at $P=0$~GPa and $z_{Se}=0.351c$ at $P=12$~GPa. The obtained band structure in Fig.~\ref{fig1}(c) shows that at $P=$12 GPa, the bands are renormalized by $Z\sim0.9$ (see inset) which allows the hole-pockets to appear on the FS. The change of FS via band renormalization can be understood this way. As deduced below, the pressure-induced renormalization affects the hopping energies, not the chemical potential. Thus at each pressure a new Fermi energy appears which allows the shifting of the band along the energy direction. This result is unexpected from the standard Fermi liquid picture.  Therefore the dominant FS instability arises along ${\bm Q}_2$ at a critical pressure, and thus $s^{\pm}$ pairing with large coupling constant commences. We also semi-quantitatively reproduce the phase diagram of the $d$ and $s^{\pm}$-pairings within the conditions for best nestings along each channels.

\section{Tight-binding modeling of pressure} To grasp further insight into how $P$ modifies electronic structure, and also to enable adding correlations for pairing symmetry calculations, we use a low-energy five-bands tight-binding (TB) formalism from Ref.~\cite{DHLee} at zero pressure, and include pressure effects by band renormalizations. The TB hopping integrals $t_i$ in a given crystal is defined as $t=\langle \Psi |V_c|\Psi\rangle$, $V_c$ is Coulomb interaction between lattice and electron. Since the TB hopping involves integration over the unit-cell volume, the simplest approximation to account for the change in TB parameters due to the change in unit cell volume is $t_i (P)/t_i(0)\propto V(P)/V_0$, where $t_i(P)$ and $V(P)$ are the TB parameters and lattice volume at any given $P$ $P$, and $t_i(0)$ and $V_0$ are their corresponding values at ambient $P$. This effective theory, which is reasonably justified by the first-principles calculations, is valid when changes in the wavefunction and Coulomb potential $V_c$ as a function of pressure are negligible. For simplicity, we take $t_i (P)/t_i(0)\sim V(P)/V_0\approx m/m^* = Z$, for all bands. From the value of $Z$ or more strictly from the volume ratio, we can obtain the value of $P$ using Birch-Murnaghan equation of state\cite{Birch}
\begin{equation}
P=\frac{3B}{2}\left( x^{7/3}-x^{5/3}\right)\left[ 1-\frac{3}{4}(4-B^{\prime})(x^{2/3}-1) \right],
\end{equation}
where $x=V_0/V=1/Z$, and $B$, $B^{\prime}$ are the bulk modulus and its derivative with respect to pressure. By fitting the calculated optimal $T_c$ of the second dome to the experimental value of $\sim 48$~K, we find $B\approx B^{\prime}P=$67 GPa, which is close to the first-principle value of $B$=64-70 GPa for this sample, and also to the available experimental data of $B$=62 GPa for other iron-pnictides.\cite{bulkmodulus}

In what follows, the renormalization, applied to the TB parameters, reflects in the band structure as $E^i_{\bm k}=Z\xi^i_{\bm k}-E_F$, where $\xi^i_{\bm k}$ is the $i^{\rm th}$ TB band taken from Ref.~\cite{DHLee}. Unlike in FL-theory, $Z$ does not renormalize the spectral weight, and thus at each pressure a new Fermi level occurs to keep the number of electron unchanged. This constraint allows the change in FS upon renormalization of the bands. We compute the Fermi level $E_F$ self-consistently, beyond a simple rigid band shift approximation, by integrating the density of state upto $E_F$. This electron number can be related to the FS volume in a slightly revised Luttinger theory as
\begin{equation}
n=\frac{1}{N}\sum_{{\bm k},i=1-5}\int \frac{d\omega}{\pi} \delta(\omega-ZE_{\bm k}^i) = \frac{1}{N}\sum_{{\bm k},i=1-5}\frac{\delta(E_{\bm k}^i)}{Z}=\frac{V_L(0)}{Z},
\end{equation}
where $V_L(0)$ is the unitless FS volume at $P=0$. This formula can be contrasted with the conventional Luttinger theorem which says that the FS volume, after applying a FL-like renormalization, does not change, because $n=\frac{1}{N}\sum_{{\bm k},i=1-5}\int \frac{d\omega}{\pi} Z\delta(\omega-ZE_{\bm k}^i) = \frac{1}{N}\sum_{{\bm k},i=1-5}Z\frac{\delta(E_{\bm k}^i)}{Z}=V_L(0)$, where $V_L$ should be read as bare FS volume. The important different between the present formalism and FL thus comes from the fact that while in FL $Z$ renormalizes both spectral weight and dispersion, in the present case it only renormalizes the band not the spectral weight. This allows the change in FS topology and volume as a function of pressure, while $V_L/Z$ remains constant, constrained by the number of of electron.
\begin{figure}[top]
\hspace{1cm}
\rotatebox[origin=c]{0}{\includegraphics[width=0.9\columnwidth]{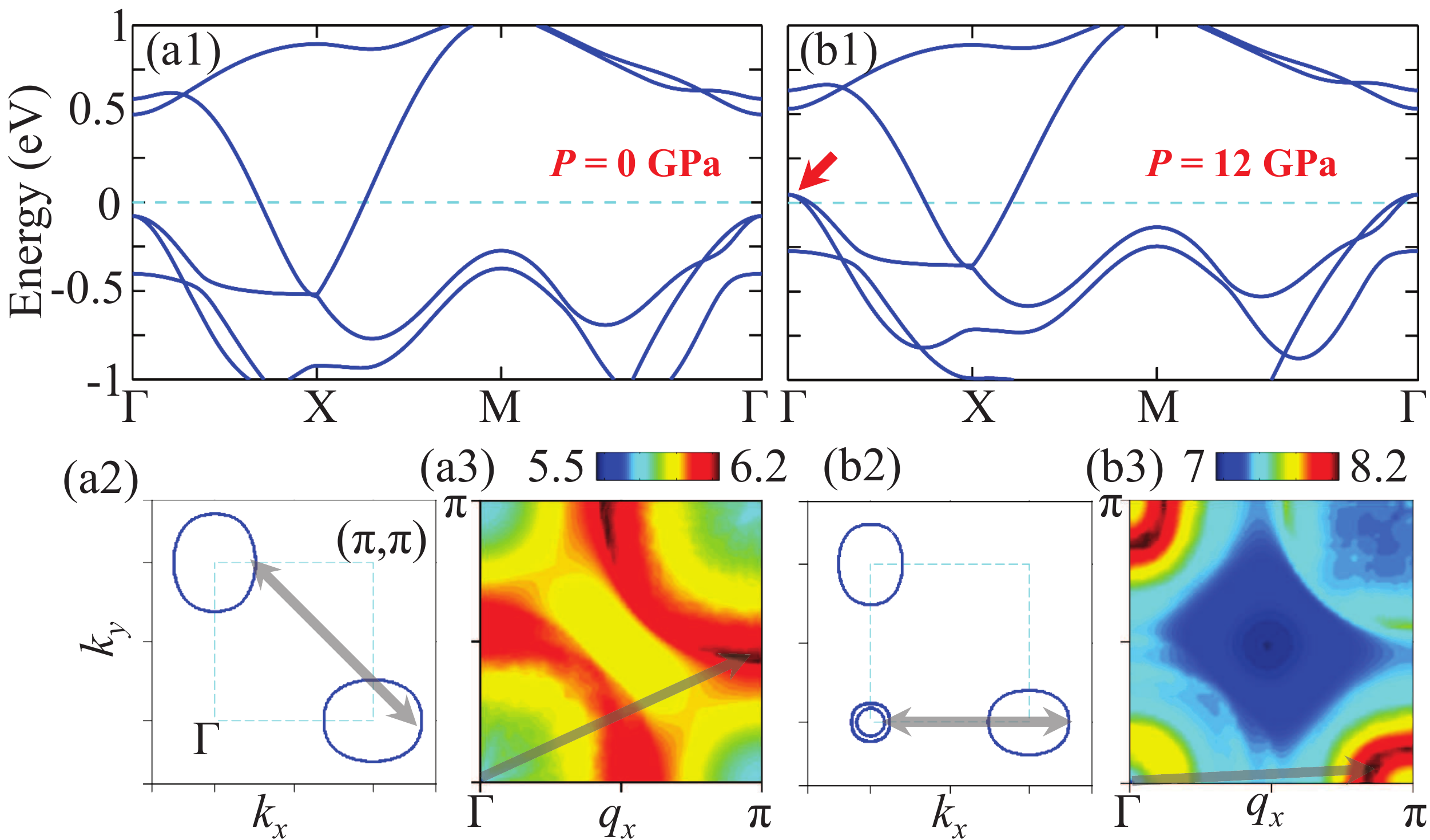}}
\caption{(a1-a3) TB bands and real-part of susceptibility at zero energy at ambient $P$. (b1-b3) Same but at $P=12$~GPa and $Z=0.9$. (a1), (b1) Self-consistently evaluated bands in 1 Fe unit cell notation\cite{DHLee} at two $P$ values. (a2), (b2) Corresponding FS topologies. (a3), (b3) Static susceptibilities plotted in the  ${\bm q}=0$-plane. Arrows depict the leading nesting directions.} \label{fig2}
\end{figure}

Figure~\ref{fig2} reveals the evolution of electronic states at ambient $P$ and at $P$=12 GPa. At $P=0$, the hole pockets lies at 60~meV below $E_F$ at $\Gamma$-point, as is also seen in ARPES data.\cite{ARPES} In this case, the presence of electron pockets at X-points (in 1 Fe per unit cell notation) is well established.\cite{ARPES,HDing,XJZhou} With the uniform band renormalization by $Z=0.9$ at $P=12$GPa, according to Eq.~1, we find that two concentric hole pockets are fully formed on the FS as seen from Figs.~\ref{fig2}(b1) and  \ref{fig2}(b2).

To determine the dominant nesting `hot-spot', we calculate multiband susceptibility using full orbital overlap matrix-element. The many-body correction is incorporated within RPA via explicitly including intraorbital interaction $U$, interorbital interaction $V=U-2J$, Hund's coupling $J=U/4$ and the pair hopping energy $J^{\prime}=J$ as defined in Ref.~\cite{footU}, and the details of the calculation can, for example, be found in Refs.~\cite{Maier,Das,paireigen,Ueda}. We present results as a function of $U$ as given in Fig.~4(a), where the other parameters change accordingly following these equations.  We plot the static spin RPA susceptibility (trace of the susceptibility tensor) at the two representative $P$s in Fig.~\ref{fig2}(a3) for $P=0$ and in Fig.~\ref{fig2}(b3) for $P=12$~GPa. As expected, the dominant nesting in the former case is aligned along the inter-electron-pocket direction [consistent with earlier calculation in Refs.~\cite{Maier,DHLee,Das}]. On the other hand at $P=12$GPa, the dominant nesting changes to ${\bm Q}_2\rightarrow (\pi,0)$ between electron and hole pockets, as obtained for many iron-pnictide superconductors.\cite{Mazin,paireigen}

\section{Pairing eigenvalue calculation}
\begin{figure}[top]
\hspace{-0cm}
\rotatebox[origin=c]{0}{\includegraphics[width=0.99\columnwidth]{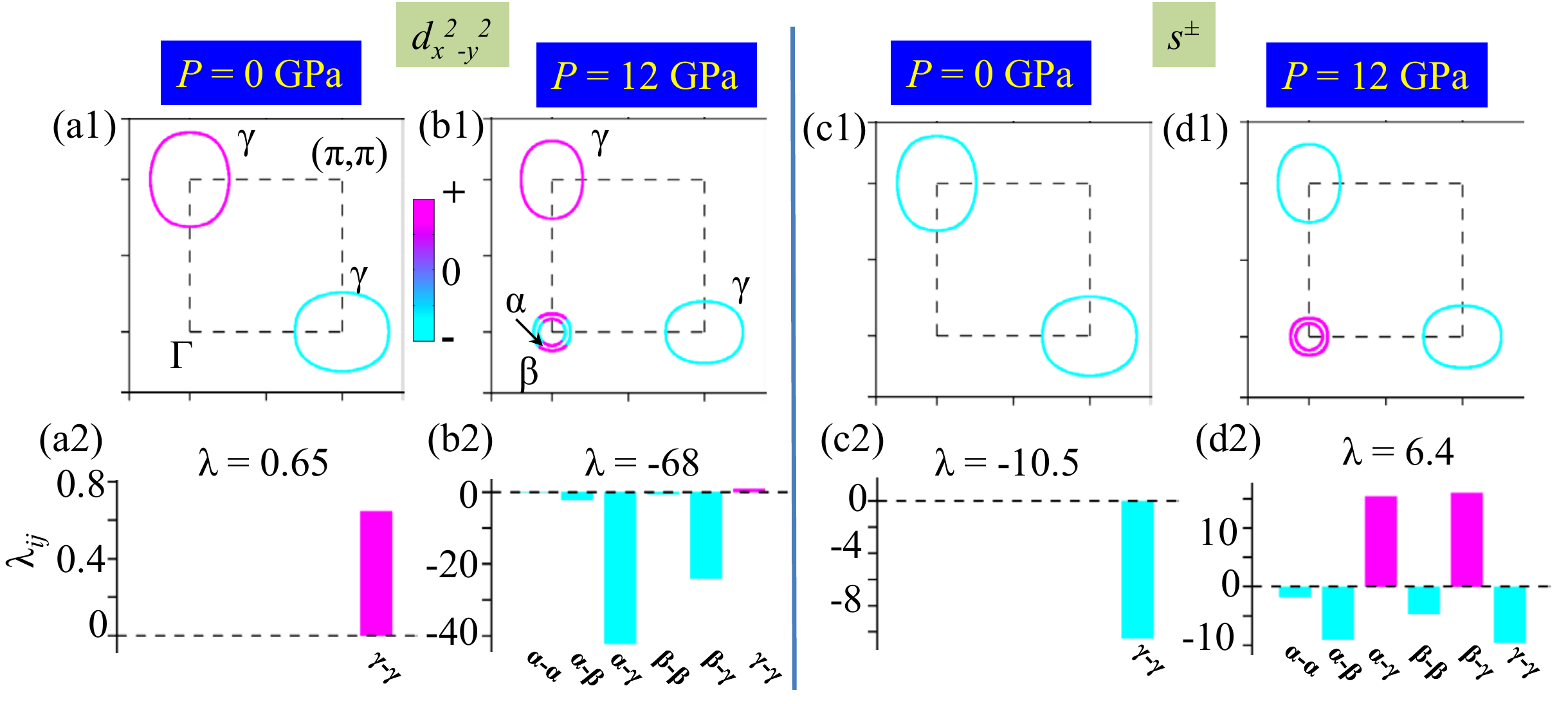}}
\caption{Evolution of different intra- and inter-band pairing eigenvalues, $\lambda_{ij}$, at two representative $P$s (different columns) and pairing symmetries (different rows). Color coding on the FS in (a1), (b1), (c1), and (d1) gives the variation and sign of SC gaps on each FS pockets. At $P=0$ (left column), the pairing eigenvalue arising from nesting between two electron pockets is positive for $d$-wave, and negative for $s^{\pm}$pairing, indicating that the former pairing is stable here. At $P=12$~GPa, the strong negative value of $\lambda_{\alpha-\gamma}$ and $\lambda_{\beta-\gamma}$, governed by same sign of SC gap for $d$-wave pairing connected by dominant `hot-spot', yields total $\lambda<0$, and thus unstable pairing. For $s^{\pm}$-pairing at this $P$, despite the presence of several negative values of $\lambda$, the dominant `hot-spot' connects sign-reversal SC gap between two electron and two hole pockets, leading to total $\lambda>0$. If one of the hole-pocket disappears, $s^{\pm}$-pairing will become unlikely in this case.} \label{fig3}
\end{figure}

\begin{figure}[top]
\hspace{3cm}
\rotatebox[origin=c]{0}{\includegraphics[width=0.7\columnwidth]{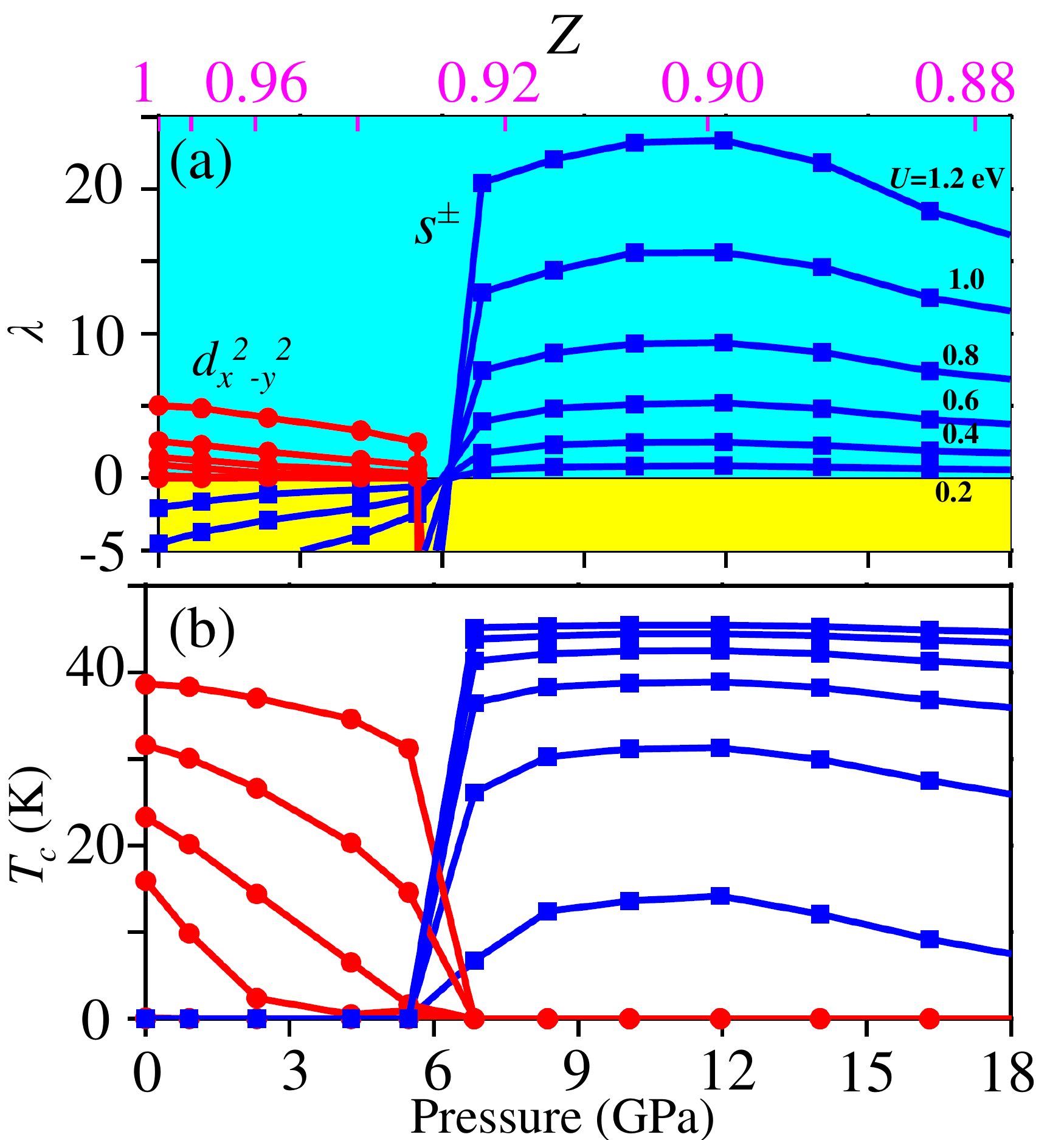}}
\caption{(a) Evolution of total pairing eigenvalues for $d$-wave (solid circles) and $s^{\pm}$-pairing as a function of $P$ for various values of $U$. The horizontal axis at the top of the plot gives the computed value of $Z$ from Eq.~1. The color shadings separates the negative and positing eigenvalue regions. (b) Computed values of $T_c$ (see text), are plotted as a function of $P$. The persistent of $T_c$ in the second SC region upto high $P$ is observed in a number of materials other than K$_0.8$Fe$_{1.7}$Se$_2$.\cite{Newdata}} \label{fig4}
\end{figure}

Next we evaluate the effective pairing vertex in the singlet channel for scattering between two FSs $i,j$ within spin and charge fluctuations exchange approximation
\begin{eqnarray}
{\hat \Gamma}({\bm k},{\bm k}^{\prime})&=&\frac{1}{2}\left[3{\hat U}_s{\hat \chi}_s({\bm k}-{\bm k}^{\prime}){\hat U}_s - {\hat U}_c{\hat \chi}_c({\bm k}-{\bm k}^{\prime}){\hat U}_c \right.\nonumber\\
&& \left.+ \frac{1}{2}({\hat U}_s+{\hat U}_c){\hat \chi}_0({\bm k}-{\bm k}^{\prime})({\hat U}_s+{\hat U}_c) \right].
\label{gamma}
\end{eqnarray}
It should be noted that the above equation is similar to the one used in Refs.~\cite{paireigen,Ueda,Yao,Schmalian,Kontani} with differences being in the bare bubble and onsite Coulomb repulsion terms only. However, since the dominant contributions to the pairing channel come from the RPA term,  the absence or presence of the other comparatively weaker terms does not alter the resulting pairing symmetry, except giving an overall shift in the value of the pairing strength. We now study the evolution of the pairing strength $\lambda$ for a given gap function $g({\bm k})$ from the following equation\cite{paireigen}:
\begin{eqnarray}
\lambda_{ij} [g] = - \frac{\oint_{c_i} \frac{d k}{v_F (k)} \oint_{c_j} \frac{d k^{\prime}}{v_F (k^{\prime})} g({\bm k}){\rm Re}\big[\Gamma_{ij}({\bm k},{\bm k}^{\prime})g({\bm k}^{\prime})\big]}{(2\pi)^2\oint_{c_j} \frac{d k}{v_F (k)} [g({\bm k})]^2}.
\label{gammaband}
\end{eqnarray}
Here the line integrals over $C_i$ are performed on each FS loops, and $v_F$ is the Fermi velocity. The key point of Eq.~\ref{gammaband} is that if the gap function $g({\bm k})$ possesses opposite sign at ${\bm k}$ and ${\bm k}^{\prime}$, mediated by a large peak in $\Gamma$ at the `hot-spot' ${\bm q}={\bm k}-{\bm k}^{\prime}$. Based on this framework, we now study the evolution of $\lambda$ for $g({\bm k})=\cos{k_x}-\cos{k_y}$ for $d_{x^2-y^2}$ and  $g({\bm k})=\cos{k_x}\cos{k_y}$ for $s^{\pm}$-pairing channels (in the 1 Fe per unit cell notation) as a function of $P$ in Figs.~\ref{fig3} and \ref{fig4} (in 2 Fe per unit cell, the form of the pairing structure $g({\bm k})$ transforms by the same unitary transformation as the FS so that the macroscopic properties such as nodeless and isotropic gap structure remains same in any notation\cite{Das,Dasmodulated}). To understand the origin of the pairing symmetry transition in details, we study each intra- and inter-band components of $\lambda_{ij}$ in Fig.~\ref{fig3} at two representative $P$s. These results confirm our initial assumption.

For $d_{x^2-y^2}$-wave pairing at $P=0$ in Fig.~\ref{fig3}(a) and at $P=12$~GPa in Fig.~\ref{fig3}(b), we obtain $\lambda>0$ and $<0$, respectively. This can be understood from the corresponding FS topological changes. At $P=0$, the two electron pockets at $(\pi,0)$ and $(0,\pi)$ posses opposite sign of SC gap, and this phase is supported by nesting along ${\bm Q}_1\rightarrow (\pi,\pi)$, as deduced in Fig.~\ref{fig2}(a3).\cite{Maier,DHLee,Das} So we get $\lambda_{\gamma-\gamma}>0$. But as the dominant `hot-spot' changes to  ${\bm Q}_2\rightarrow (0,\pi)$ at $P=12$~GPa, positive $\lambda_{\gamma-\gamma}$ component is overturned by large negative $\lambda_{\alpha-\gamma}$ and $\lambda_{\beta-\gamma}$ components being supported by ${\bm Q}_2$ `hot-spot' without sign-reversal of $g({\bm k})$. Therefore, the total $\lambda$ becomes negative, making $d_{x^2-y^2}$ pairing unstable at this $P$. For $s^{\pm}-$pairing, the situation is reversed in that $\lambda_{\gamma-\gamma}<0$ at all values of $P$, but $\lambda_{\alpha/\beta-\gamma}>0$. It is important to note that due to the presence of two hole pockets $\alpha$ and $\beta$, the total value of $\lambda$ becomes positive and large at some critical value of $P$, otherwise, $s^{\pm}$-pairing would have been favorable.

In Fig.~\ref{fig4}, we show the full $P$ and interaction $U$ dependence of the total pairing strengh $\lambda$, and the corresponding calculated $T_c$. The $d$-wave pairing survives upto $P\sim5$~GPa, slightly less than the termination point of the first SC shown in Fig.~\ref{fig1}(a). Of course, these quantitative consistencies rely strongly on the exact shape of FS topology and the value of bulk modulus $B$ used in Eq.~1. We focus on fitting $B$ so that the optimum $T_c$ for the second dome match with the experimental value of $P$.\cite{KFe2Se2P,Newdata} With a separation of about 1-2 GPa, we find that the $s^{\pm}-$pairing channel appears abruptly for a large range of $U$. Although we obtain an optimum $\lambda$ as a function of $P$ at which the FS nesting between $\alpha/\beta$ to $\gamma$ is strong for all values of $U$ considered, it survives to a larger $P$ range than the experimental data of K$_{0.8}$Fe$_{1.7}$Se$_2$.\cite{KFe2Se2P} However, for other samples within the same family, new data shows that $T_c$ in the second dome is very much $P$ independent and survives up to $P$ as large as 40~GPa measured so far.\cite{Newdata}

For a purely electronic mechanism, we compute $T_c$ using spin-fluctuation exchange formula within the weak-coupling limit\cite{Carbotte,Dynes,Pines,Schrieffer,Markiewicz}
\begin{eqnarray}
T_c=\frac{\omega_{sf}}{1.2}\exp{(-1.04/\lambda)},
\end{eqnarray}
with $\omega_{sf}$=55~K, we obtain optimum $T_c\sim$38~K and $T_c\sim45$~K for the first and second dome, respectively, for $U=1.2$~eV, which are close to the experimental values of 37~K and 48~K. It is obvious that the absolute value of $T_c$ depends on the interaction parameter, however, the ratio between the optimum value at two domes is always maintained. Interestingly, we find that the computed $T_c$ for the second dome is even flatter than that for $\lambda$ and thus agrees well with the new data.\cite{Newdata}

\begin{figure}[top]
\hspace{3cm}
\rotatebox[origin=c]{0}{\includegraphics[width=0.6\columnwidth]{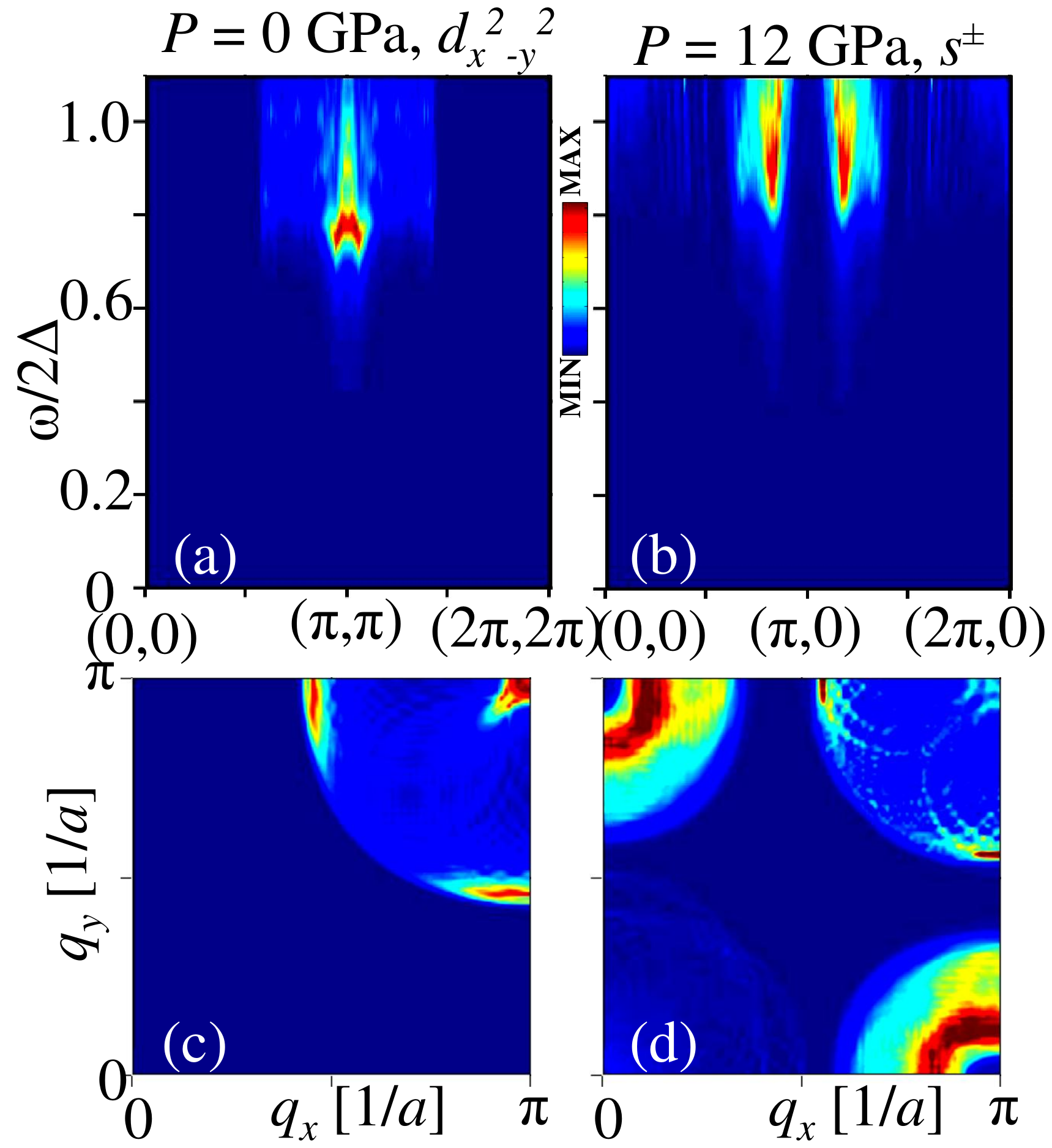}}
\caption{(a) Spin excitation spectrum along zone diagonal direction as a function of energy for $P=0$, and $d$-wave pairing. (b) Same but along zone boundary direction at $P=12$~GPa and $s^{\pm}$-pairing. (c)-(d) Constant energy cuts at their corresponding resonance energy values. } \label{fig5}
\end{figure}

\section{Pressure dependence spin-resonance}

Relating the glue function to the spin-resonance mode that appears in the SC state, we obtain a resonance $\omega_{res}\sim15$~meV at the optimum $P$. The phenomena of spin-resonance in the SC state is well known\cite{Scalapino,Chubukov,Mazin,Maier,paireigen} and is essentially similar to the condition for obtaining positive pairing eigenvalue $\lambda$ as discussed above: Given that the sign reversal of SC gap is connected by momentum transfer ${\bm q}$ a spin-resonance appears at an energy $\omega({\bm q})\sim |\Delta({\bm k})|+|\Delta({\bm k}+{\bm q})|$ (RPA correction shifts the mode to a slightly lower energy). We compute the spin-resonance spectra at the same representative $P$ values as in Figs.~\ref{fig2} and \ref{fig3} for $d$-wave and $s^{\pm}$-pairing and the results are shown in Fig.~\ref{fig5}. We immediately see that for $d$-wave pairing we get a nearly commensurate spin-resonance mode at $\omega/2\Delta\sim0.8$, while it becomes incommensurate and shifts to higher energy for the $s^{\pm}$-pairing at higher $P$. Although, ARPES and INS measurements are difficult to perform under $P$, neutron diffraction experiment can be done here to test the changes of ${\bm Q}$ vector as a function of $P$ to verify our results.

\section{Discussion and conclusions}

The FS topology modeled here is constrained by the consistency between band-structure calculations and ARPES measurements, and thus we expect that correlation effects driving either coexistence with vacancy order and/or magnetic phase,\cite{WBao,Dasmodulated} or phase separation between them\cite{phaseseparation} will not dramatically change our results but more studies are needed to address this question. Furthermore, we also note that a recent ARPES measurement\cite{nodelessARPES} has shown that a tiny hole pocket develops around $Z$-point at ambient $P$ with an isotropic SC gap. In our present scenario of $d$-wave pairing , one would expect a node on this FS. However, for such tiny FS elevated along the $k_z$ direction and small nodal quasiparticle weight, one can expect the node to become filled due to various extraneous effects such as disorder, final state scattering of ARPES measurement. A  recent experimental study on LaFeAsO$_{1-2}$H$_x$ found two SC dome with a dip in $T_c$ as a function of doping $x$.\cite{Hosono}. Based on first-principle band-structure calculation, it has been argued that the pairing changes  from $s^{\pm}$ to $s^{++}$  in that case due to orbital degeneracy.

In conclusion, we present an analysis of pressure evolution of pairing interaction in $A_y$Fe$_{2-x}$Se$_2$ family of superconductors. We argue that observed two SC domes can be naturally explained by changes in FS topology driven by mass enhancements. At lower $P$, we expect no $\Gamma$-point hole pocket on the FS, and therefore $d$-wave SC state.\cite{nodelessARPES} Upon increase in $P$, we expect two hole pockets to develops at $\Gamma$-point, leading to pairing symmetry transition to $s^{\pm}$-pairing. To test these predictions, we suggest to use magnetic field dependent tunneling spectroscopies to investigate gap changes.\cite{Hanaguri} Neutron scattering and neutron diffraction measurements would reveal the distinct pattern of spin resonance and nesting properties in these phases.

\section*{Acknowledgments}
This work was supported, in part, by UCOP-TR,  by Nordita and by Los Alamos National Laboratory, of the US Department of
Energy under Contract DE-AC52-06NA25396, and benefited from the allocation of supercomputer time at NERSC.

\section{Appendix}

The spin-fluctuation mediated pairing interaction is well studied in the literature\cite{Scalapino,paireigen,Ueda,Yao,Schmalian,Kontani}, which sometimes include the bare bubble term in Eq.~\ref{gamma} and/ or the onsite interaction term (not included here).  The spin-fluctuation spectrum in the unconventional superconductors (SCs) obtains a sharp peak at a resonance energy, $\omega_{res}$, and to then it falls off sharply on the energy scale (see Fig.~5 of main text). Therefore, although the SC gap equation depends on the full vertex, Im~$\big[{\hat \Gamma}({\bm k},{\bm k}^{\prime},\omega)\big]$, the relevant ${\bm k}$ and ${\bm k}^{\prime}$ values are restricted by this energy cutoff to remain in the vicinity of the Fermi surfaces (FSs). In this spirit, just as for the electron-phonon case, the strength of the pairing interaction is characterized by an energy integral over ${\hat \Gamma}$ weighted by $\omega^{-1}$ following the Kramers-Kronig relation as:
\begin{eqnarray}
\int_0^{\infty}\frac{{\rm Im}\big[{\hat \Gamma}({\bm k},{\bm k}^{\prime},\omega)\big]}{\pi\omega}d\omega={\rm Re}\big[{\hat \Gamma}({\bm k},{\bm k}^{\prime},\omega=0)\big].
\label{KKR}
\end{eqnarray}
This allows us to evaluate the pairing strength by considering only the real part of the static pairing interaction. This leads to solving the pairing eigenvalue problem by integrating over a closed FS for a given pairing symmetry, say $g({\bm k})$, as
\begin{eqnarray}
-\sum_i\oint_{C_i}\frac{dk}{2\pi}\frac{{\rm Re}\big[\Gamma_{ij}({\bm k},{\bm k}^{\prime})\big]}{2\pi v_F}g({\bm k})=\lambda g({\bm k}^{\prime}),
\label{eigenvalue}
\end{eqnarray}
where $v_F$ is the Fermi velocity and $C_i$ gives the closed FS for the $i^{th}$-band. If the gap function can be decomposed by its amplitude $\Delta_0$ and structure factor $g({\bm k})$, then the above eigenvalue problem can be reduced to a dimensionless pairing strength functional\cite{Scalapino} given in Eq.~\ref{gammaband}.
%
%
The total pairing strength is then obtained by summing over all bands.
%

Combining Eq.~\ref{gammaband} and Eq.~\ref{KKR}, we can recast the eigenvalue problem in the typical Eliasberg framework as
\begin{eqnarray}
\lambda[g] = \frac{2}{\pi}\int_0^{\infty}\frac{\alpha^2F(\omega)[g]}{\omega}d\omega,
\label{lambda1}
\end{eqnarray}
where the electron-boson spectral function $\alpha^2F$ is nothing but a momentum average over the dynamical pairing interaction weighted by the gap function $g$ as
\begin{eqnarray}
\alpha^2F(\omega)[g] = - \frac{\oint_{c_i} \frac{d k}{v_F (k)} \oint_{c_j} \frac{d k^{\prime}}{v_F (k^{\prime})} g({\bm k}){\rm Im}\big[\Gamma_{ij}({\bm k},{\bm k}^{\prime},\omega)\big]g({\bm k}^{\prime})}{(2\pi)^2\oint_{c_j} \frac{d k}{v_F (k)} [g({\bm k})]^2}.
\end{eqnarray}

For any general electron-boson interaction (including phonon\cite{Dynes} and spin-fluctuation\cite{Carbotte,Pines,Schrieffer,Markiewicz}), SC transition temperature, $T_c$ can be calculated from the pairing strength $\lambda$ in the weak-coupling scenario as
\begin{eqnarray}
T_c=\frac{\omega_{sf}}{1.2}\exp{(-1.04/\lambda)}.
\label{Tc}
\end{eqnarray}

Here the spin-fluctuation cutoff frequency $\omega_{sf}$ is given by\cite{Dynes}
\begin{eqnarray}
\omega_{sf} = \exp{\left(\frac{2}{\lambda}\int_0^{\infty}\log{\omega}\frac{\alpha^2F(\omega)}{\omega}d\omega\right)}.
\label{wsf}
\end{eqnarray}
Finally, as discussed before, we assume that the $\alpha^2F$ has a strong peak at $\omega_{res}$, and falls off rapidly away from this energy. Using Eq.~\ref{lambda1}, we obtain $\omega_{sf}\approx\omega_{res}$. In Fig. 4 of the main text, we evaluate $T_c$ by using Eqs.~\ref{Tc} and \ref{wsf} with the coupling constant evaluated from Eq.~\ref{gammaband}.


\begin{thebibliography}{10}
\bibitem{footU} The spin and charge RPA susceptibilities are defined as ${\hat \chi}_s = {\hat \chi}_0/(1-{\hat U}_s{\hat \chi}_0)$ and ${\hat \chi}_c = {\hat \chi}_0/(1+{\hat U}_s{\hat \chi}_0)$, where ${\hat \chi}_0$ is the non-interacting susceptibility tensor, and ${\hat U}_{s/c}$ are spin and charge interaction vertex, defined in the orbital basis, see Ref.~\cite{Ueda}.
\end{thebibliography}
\end{document}